\documentclass[aps,twocolumn]{revtex4}
\newcommand{\Rv}{\mbox{{\bf R}}}

\newcommand{\Av}{\mbox{{\bf A}}}
\newcommand{\Qv}{\mbox{{\bf Q}}}
\newcommand{\lv}{\mbox{{\bf l}}}
\usepackage{epsf}
\usepackage{graphicx}
\begin{document}
\title{Critical field in a model with local pairs}
\author{Marcin Mierzejewski and  Maciej M. Ma\'ska}
\email{maciek@phys.us.edu.pl}
\affiliation{
Department of Theoretical Physics, Institute of Physics,
University of Silesia, 40-007 Katowice,
Poland}
\begin{abstract}
We analyze the role of Zeeman and orbital pair breaking mechanisms
in models appropriate for short coherence length
superconductors. In particular, we investigate 
the attractive Hubbard  and the pair hoping
models. The orbital pair breaking mechanism dominates in the majority of models with
$s$--wave and $d$--wave superconducting order parameters. On the other, the
repulsive pair hopping interaction leads to  $\eta$--type pairing, that is stable
against the orbital pair breaking. External magnetic field reduces 
this type of pairing predominantly due to the Zeeman coupling.
According to the recent experiments this mechanism is responsible for
closing of the pseudogap. Moreover, the temperature dependence 
of the gap closing field in $\eta$--phase fits the experimental
data very well.  We discuss whether the preformed pairs 
in the $\eta$--phase could be responsible for the  pseudogap phenomenon.
\end{abstract}
\pacs \ 74.25.Ha,74.20.-z,74.20.Rp
\maketitle
\section{Introduction}
Almost twenty years after discovery of high--temperature superconductors (HTSCs)
the mechanism responsible for their unusual properties  
remains unclear. The complex phase diagram of HTSCs suggests
that there may be no single mechanism that dominates over the
entire doping range. In particular, the normal--state properties 
in underdoped and overdoped regimes are different. Highly overdoped compounds 
in the normal 
state exhibit Fermi liquid behavior, whereas the superconducting state
may be described within a weak--coupling BCS theory\cite{Fermi}. 
On the other hand, in the underdoped regime the HTSCs exhibit unconventional
features. The most remarkable of them are the extremely short coherence length and a 
pseudogap that opens in the normal state. The presence of the pseudogap has 
been confirmed with the help of various experimental techniques like:
angle--resolved photoemission \cite{arpes1,arpes2,arpes3},
intrinsic tunneling spectroscopy \cite{krasnov1,krasnov2},
NMR \cite{nmr1,nmr2}, infrared \cite{infra} and transport \cite{transport} 
measurements. Although, there is no complete theoretical description of the
pseudogap, one usually considers this phase as a precursor of the superconductivity. 
According to this hypothesis formation of Cooper pairs starts
at temperature $T^*$, higher than the superconducting transition temperature $T_c$.
Then, at $T_c$, these preformed pairs undergo Bose-Einstein condensation.

This hypothesis seems to be supported by recent observations of the vortex--like
Nernst signal above $T_c$ \cite{xu} that evolves smoothly into the analogous signal
below the superconducting phase transition \cite{ochida}. The Meissner effect
does not occur in the pseudogap phase due to strong phase fluctuations rather
than the vanishing of the superfluid density. Therefore, theoretical description of
the suppression of the Meissner effect requires an approach beyond 
the mean--field level. Despite the absence of the Meissner effect above $T_{c}$,
one can observe inhomogeneous magnetic domains that are interpreted as
precursors to the Meissner state \cite{iguchi}. 

The short coherence length indicates that the pairing takes place in the real space,
leading to boson--like objects. A few models are commonly used to describe 
systems with the local pairs. 
Namely, the attractive Hubbard (AH) model \cite{micnas},
fermion--boson\cite{FB} and purely bosonic models\cite{boson}, as well as
the Penson--Kolb (PK) model\cite{PK}, i.e., the tight--binging model with local pair
hopping. These models should be considered
as effective approaches which do not explain the microscopic origin of the
pairing interaction. 

Another unusual property of HTSCs is related to their behavior in the external 
magnetic field. In particular, temperature dependence of the upper
critical field $H_{c2}$ has a positive curvature\cite{osofsky,mackenzie} 
in contradistinction to classical
superconductors, where a negative curvature is observed. 
Moreover,  $H_{c2}$ does not saturate even at genuinely low temperature.
Recent experiments\cite{hpg} show that also the pseudogap is destroyed by sufficiently
high magnetic field, $H_{pg}$. Although, the temperature dependence of $H_{pg}$ has
a negative curvature, it significantly
differs from the predictions of the standard Helfand--Werthammer theory\cite{HW}.
Namely, $H_{pg}(T)$ has a large slope at temperatures close to $T^*$ 
and saturates already at $T\simeq 0.7 T^*$.
These features may assist in verification of the 
preformed Cooper pairs hypothesis and, more generally, in choosing the most 
appropriate model of HTSC.
   
\section{Model}

In the present paper we show that opening of the pseudogap and its dependence 
on the magnetic field can be described within a model with local pair hopping. 
Our starting point is the two--dimensional (2D) Penson--Kolb model with the 
Hamiltonian given by:
\begin{eqnarray}
H &= &\sum_{i,j,\sigma} t_{ij} {\rm e}^{i \Phi_{ij}} 
c^{\dagger}_{i \sigma} c_{j \sigma} + \sum_{i \sigma}
(g \mu_B H_z \sigma-\mu) 
c^{\dagger}_{i \sigma} c_{i \sigma} \nonumber \\
&& -\frac{1}{2}J \sum_{\left<i,j\right>} {\rm e}^{2 i \Phi_{ij}}
c^{\dagger}_{i \uparrow}c^{\dagger}_{i \downarrow}  c_{j \downarrow}
c_{j \uparrow}.
\end{eqnarray}
Here, $c^{\left(\dagger \right)}_{i \sigma}$ creates (annihilates)
an electron with spin $\sigma$ at site $i$, $t_{ij}$ is the single
electron hopping integral between sites $i$ and $j$,
$\mu$ is the chemical potential and
$J$ is the nearest neighbor pair hopping interactions.
The external magnetic field perpendicular to the lattice 
$H_z$ shifts the energy levels
by $g \mu_B H_z \sigma$ ($g$ is the gyromagnetic ratio
and $\mu_B$ is the Bohr magneton) and modifies the
hopping terms. The single electron hopping  
integral acquires the Peierls factor 
\begin{equation}
\Phi_{ij}= \frac{e}{\hbar c} \int^{\Rv_{i}}_{\Rv_{j}}
\Av\cdot d\lv, 
\end{equation}
whereas the phase factor in the pair hopping term
is twice larger.

The Penson--Kolb model can be derived from a general microscopic 
tight--binding Hamiltonian\cite{hubbard}, where the Coulomb repulsion
may lead to the pair hopping interaction. In such a case $J$ is negative 
(repulsive Penson--Kolb model). However, treating
the Penson--Kolb model as of a phenomenological nature, we assume
$J$ to be an effective parameter, that can be negative as well as
positive. It can be understood as a result of renormalization originating,
e.g., from  electron--phonon coupling\cite{rob-bul}.
For a nonzero single electron hopping integral $J \rightarrow -J$ is not a symmetry
of the PK model \cite{japaridze}.
However, superconducting correlations occur in the Penson--Kolb model for attractive
pair hopping interaction ($J>0$) as well as
for the repulsive one ($J<0)$, provided that the pair
hopping is large enough. The latter case is
usually referred to as $\eta$--type pairing. Then,
the total momentum  of the paired electrons is $\Qv=(\pi,\pi)$
and the phase of superconducting order parameter alters
from one site to the neighboring one. It has been shown
that there is a flux quantization and Meissner effect
in this state.\cite{yang} 
Superconductivity survives also in the presence of on--site Coulomb 
repulsion (Penson--Kolb--Hubbard model), provided that this interaction 
is not too large\cite{Fab}. 

\subsection{Density of states}

At the mean--field level, for $J>0$ one obtains an isotropic
superconducting gap, identical to that obtained for AH model.
On the other hand, in the case of $\eta$--type pairing ($J<0$),
the density of states is finite for arbitrary energy. However,
the density of states at the Fermi level may significantly be suppressed
for some dopings. In the simplest case of the nearest neighbor hopping the
density of states in the $\eta$--phase is of the form:
\begin{equation}
\rho(\omega)=\frac{1}{2}\left(1-\frac{\mu}{\tilde{\mu}}\right)
\rho_0(\omega-\tilde{\mu})
+
\frac{1}{2}\left(1+\frac{\mu}{\tilde{\mu}}\right)
\rho_0(\omega+\tilde{\mu}) \label{rho}.
\end{equation}
Here, $\tilde{\mu}=\sqrt{\mu^2+4 |J\Delta|^2}$, 
$\Delta \equiv (-1)^i \left< c_{i \downarrow} c_{i \uparrow} \right>$ 
is the $\eta$--phase order parameter and $\rho_0$ is the density of states
for $\mu=\Delta=0$. One can see from Eq. (\ref{rho}) that the
quasiparticle poles split when $\Delta$ becomes finite. Therefore, a
local minimum in the density of states may occur at the Fermi surface. 
Despite the presence of this minimum the density of states at the
Fermi level remains finite provided that $\tilde{\mu}$ is small
when compared to the band width.
Inclusion of the next nearest neighbor hopping $t'$ leads to a
more complicated expression for the density of states.
However, the structure of $\rho$ in the
$\eta$--phase remains unchanged. Fig. 1. shows the density of states
calculated for $t'\neq 0$ and different values of the $\eta$--phase
order parameter.
\begin{figure}
\centering
\includegraphics[width=7.5cm]{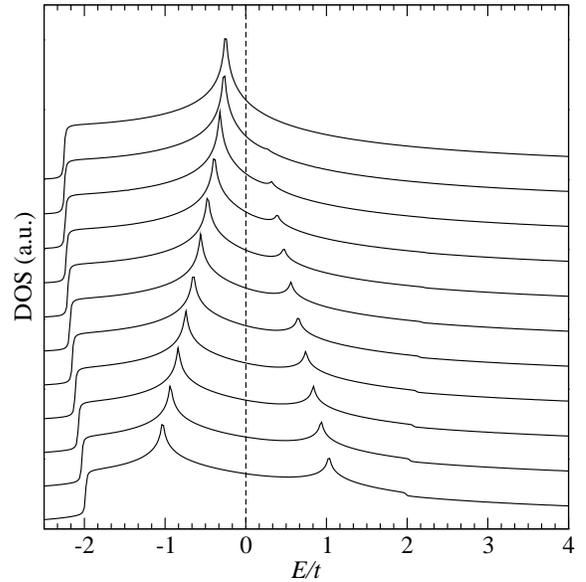}
\caption{Density of states for the Penson--Kolb model in the $\eta$--phase.
We have used $t'=-0.25\:t,\ J=-t$ and $\mu=-0.75\: t$. The curves from the topmost
to the lowest correspond to the values of the order parameter
$\Delta=0,\ 0.1,...,1.$ The dashed line indicates the Fermi level.}
\label{fig_dos} 
\end{figure}
Gradual decreasing of $\rho$ at the Fermi level resembles
opening of the pseudogap in HTSCs. 

Another feature that could speak in favor of this interpretation 
is anisotropy of the gap \cite{arpes2}. More precisely, for $t'\neq 0$ 
the magnitude of splitting of the quasiparticle peaks 
depends on the direction in the Brillouin zone. The splitting of the
spectral functions is presented in Fig. 2. 
As we consider isotropic order parameter the splitting is finite
everywhere at the Fermi level, in contradistinction to a purely
$d$--wave gap. 
However, this drawback may be removed when considering a nonlocal pairing.
\begin{figure}
\centering
\includegraphics[width=7.5cm]{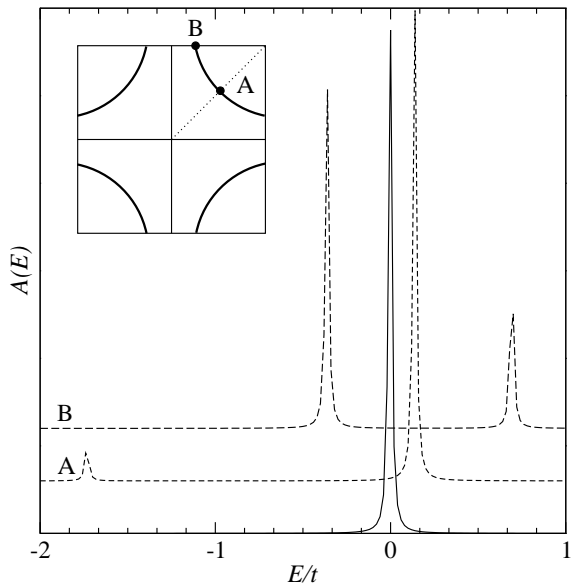}
\caption{The spectral functions in various points at the Fermi surface.
We have used the same model parameters as in Fig. 1. The position of points 
A and B is depicted in the inset. The continuous curve corresponds to the 
case $\Delta=0$, whereas the dashed lines have been obtained for $\Delta=0.5$.}
\label{arpes} 
\end{figure}

\subsection{Response to magnetic field}	
	
In contradistinction
to the AH model, the external magnetic
field explicitly enters the term  responsible
for superconductivity, i.e., the pair
hopping interaction.
Therefore, the differences between
AH and PK models may show up in the electromagnetic properties \cite{czart}.
Here, we investigate the temperature dependence of critical field $H_{\rm crit}$. 
It is defined as the highest magnitude of the magnetic field, for
that there exists a non--zero solution for the superconducting
order parameter:
\begin{equation}
\Delta_{i}=\left< c_{i \downarrow} c_{i \uparrow} \right>.
\end{equation}
As we carry out calculations 
at the mean--field level it is impossible to determine the phase coherence
of the Cooper pairs. Therefore, the physical interpretation of the critical field 
is not unique. In the overdoped regime of cuprates it can directly correspond
to the upper critical field. On the other hand, for underdoped systems, and
within the precursor scenario of pseudogap, it can be interpreted as $H_{pg}$,
i.e., the field at which the incoherent pairs appear. One can argue that 
the PK model can also be used as an effective model of other superconductors
that are characterized by a short coherence length: barium bismuthates, 
fullerides, Chevrel phases, organic superconductors and heavy fermion systems.
\cite{czart} In these cases $H_{\rm crit}$ should be considered as $H_{c2}$.

For the sake of simplicity we define:
\begin{equation}
\tilde{\Delta}_{i}=\frac{1}{2}{\sum_j}' 
{\rm e}^{2 i \Phi_{ij}} \Delta_{j} \label{td},
\end{equation}
where the prime means that the summation is carried out over the 
nearest neighbors of site $i$.
Then, the mean--field Hamiltonian takes on the following form:
\begin{eqnarray}
H &= &\sum_{i,j,\sigma} t_{ij} {\rm e}^{i \Phi_{ij}} 
c^{\dagger}_{i \sigma} c_{j \sigma} + \sum_{i \sigma}
(g \mu_B H_z \sigma-\mu) 
c^{\dagger}_{i \sigma} c_{i \sigma} \nonumber \\
&& -J \sum_{i}\left(
c^{\dagger}_{i \uparrow}c^{\dagger}_{i \downarrow}\tilde{\Delta}_{i}
+ {\rm H.c.}\right) \label{Hmf}.
\end{eqnarray}    
At the mean--field level the only difference between PK and AH 
models is the presence of $\tilde{\Delta}_{i}$ in Eq. (\ref{Hmf})
instead of $\Delta_{i} $. Therefore, 
in order to calculate the critical field one can follow an approach, 
that has previously been developed for the lattice gas 
with on--site attraction \cite{my}.
Then, one ends up with the lattice version of the Gor'kov equations:
\begin{equation}
\Delta_{i}= \frac{J}{\beta }\sum_{j,\omega_n}
\tilde{\Delta}_{j} G(i,j,\omega_n)G(i,j,-\omega_n) \label{ge}.
\end{equation}
Here, $G(i,j,\omega_n)$ is the one--electron Green's function
in the presence of a uniform and static magnetic field.
$\omega_n$ denotes the fermionic Matsubara frequency.
With the help of Eq.(\ref{td}) one can eliminate 
$\tilde{\Delta}_i$ from the Gor'kov equations. Then,
$H_{\rm crit}(T)$ can be calculated from
\begin{equation}
\Delta_{i}= \frac{J}{2 \beta }\sum_{\left<j,l\right>,\omega_n}
{\rm e}^{2 i \Phi_{lj}} \Delta_{j} G(i,l,\omega_n)G(i,l,-\omega_n)
\label{s}
\end{equation}
or \begin{equation}
\Delta_{i}= \frac{-J}{2 \beta }\sum_{\left<j,l\right>,\omega_n}
(-1)^{j+l}
{\rm e}^{2 i \Phi_{lj}} \Delta_{j} G(i,l,\omega_n)G(i,l,-\omega_n)
\label{eta}.
\end{equation}
Eqs. (\ref{s}) and (\ref{eta}) are equivalent 
since $(-1)^{j+l}=-1$, for the neighboring sites $j$ and $l$. However, 
it is more
convenient to use the first/second of them for attractive/repulsive 
pair hopping interaction. 

In the following we consider only
the nearest--neighbor one--particle hopping integral $t$ and use the Landau
gauge $\Av=H_{z}\left(0,x,0\right)$.  Then, the Harper equation
\begin{eqnarray}
g\left(\bar{p}_x,p_y,x+1\right)+
2\cos\left(h x -p_ya\right)g\left(\bar{p}_x,p_y,x\right) && \nonumber \\
+ g\left(\bar{p}_x,p_y,x-1\right)=
t^{-1}E\left(\bar{p}_x,p_y\right)
g\left(\bar{p}_x,p_y,x\right), &&
\end{eqnarray}
determines eigenvalues $E\left(\bar{p}_x,p_y\right)$ of the
one--particle hopping term. The corresponding eigenstates are enumerated
by $\bar{p}_x, p_y$ and are of the form:
\begin{equation}
U_{x,y}\left(\bar{p}_x,p_y\right)=
e^{\displaystyle ip_y ya}
g\left(\bar{p}_x,p_y,x\right). \label{Uxy}
\end{equation}
Here, $x,y$ are integers which enumerate the lattice sites
in $\hat x$ and $ \hat y$ directions, whereas
$h/(2 \pi)$ is a ratio of the flux through a lattice cell 
to one flux quantum. We refer to Ref. \cite{my} for the
details.

 The one--electron Green's function can be expressed with the help of
eigenvalues and eigenstates of the normal--state Hamiltonian. Then,
the summation over Matsubara frequencies in Eqs. (\ref{s}) and  
(\ref{eta}) can explicitly be carried out. In the Landau gauge
the presence of magnetic field does not change the plane--wave 
behavior in  $\hat y$--direction [see Eq. (\ref{Uxy})]. Therefore,
the superconducting order parameter depends only on $x$ and 
pairing of electrons takes place for the same $\hat y$--components
of their momenta, as in the absence of magnetic field, i.e., 
$(p_y, -p_y)$ for $J>0$ and $(p_y,\pi-p_y)$ for $J<0$.
Taking these features into account one can rewrite the Gor'kov
equations for the attractive:
\begin{eqnarray}
\Delta_{x^{\prime}}&=&\frac{J}{2\sqrt{N}}\sum_x \Delta_x
\sum_{p_y,\bar{p}_x,\bar{k}_x} \chi(\bar{p}_x,p_y,\bar{k}_x,-p_y)
 \nonumber \\
& & \times \;
\left[2\cos(2hx)\;g\left(\bar{p}_x,p_y,x\right)
g\left(\bar{k}_x,-p_y,x\right) \right. \nonumber \\
& & + \; g\left(\bar{p}_x,p_y,x+1\right)
g\left(\bar{k}_x,-p_y,x+1\right)  \nonumber \\
& & + \left. \; g\left(\bar{p}_x,p_y,x-1\right)
g\left(\bar{k}_x,-p_y,x-1\right) \right] \nonumber \\
& & \times \;
g\left(\bar{p}_x,p_y,x^{\prime}\right)
g\left(\bar{k}_x,-p_y,x^{\prime}\right),
\end{eqnarray} 
as well as for the repulsive pair hopping interaction:
\begin{eqnarray}
\Pi_{x^{\prime}}&=&\frac{-J}{2\sqrt{N}}\sum_x \Pi_x
\sum_{p_y,\bar{p}_x,\bar{k}_x} \chi(\bar{p}_x,p_y,\bar{k}_x,\pi-p_y) \nonumber \\
& & \times \;
\left[2\cos(2hx)\;g\left(\bar{p}_x,p_y,x\right)
g\left(\bar{k}_x,\pi-p_y,x\right) \right. \nonumber \\
& & - \; g\left(\bar{p}_x,p_y,x+1\right)
g\left(\bar{k}_x,\pi-p_y,x+1\right)  \nonumber \\
& & - \left. \; g\left(\bar{p}_x,p_y,x-1\right)
g\left(\bar{k}_x,\pi-p_y,x-1\right) \right] \nonumber \\
& & \times \;
g\left(\bar{p}_x,p_y,x^{\prime}\right)
g\left(\bar{k}_x,\pi-p_y,x^{\prime}\right).
\end{eqnarray}
Here $\Pi_x \equiv \Delta_{x,y}(-1)^y$ and the Cooper pair 
susceptibility $\chi(\bar{p}_x,p_y,\bar{k}_x,k_y)$ has the
following form:
\begin{eqnarray}
\chi(\bar{p}_x,p_y;\bar{k}_x,k_y)& =& \left[\tanh\frac{E\left(\bar{p}_{x},
p_{y}\right)-\mu-g\mu_BH_z}{2k_{B}T}
 \right. \nonumber \\
&+&\left. \tanh\frac{E\left(\bar{k}_{x},k_{y}
\right)-\mu+g\mu_BH_z }{2k_{B}T}
\right] \nonumber \\
&\times &
\left[2\left(E\left(\bar{p}_{x},p_{y}\right)+E\left(\bar{k}_{x},
k_{y}\right)-2\mu\right) \right]^{-1}. \nonumber \\
\end{eqnarray}

The above equations determine the strength of magnetic field at
which the local pairing disappears. We have carried out calculations
for $150\times 150$ cluster with periodic boundary conditions 
(bc) along the $\hat{y}$ axis. As the Landau gauge breaks the 
translation invariance along $\hat{x}$ axis we have used fixed bc in this 
direction. Our previous calculations indicate that such a size
of cluster is sufficient to obtain convergent results.\cite{my} 

\begin{figure}
\centering
\includegraphics[width=7.5cm]{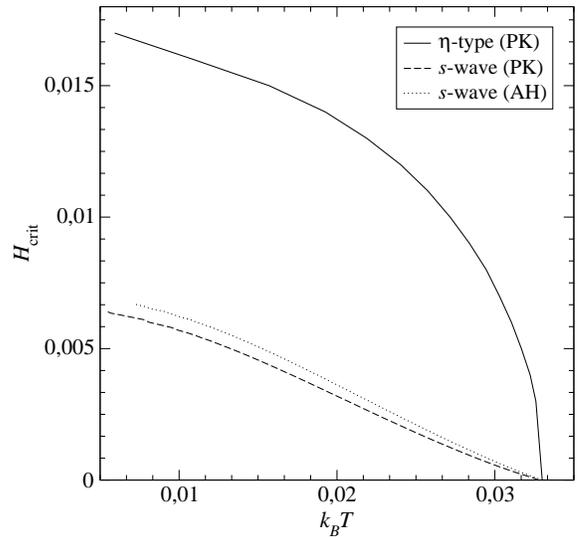}
\caption{Temperature dependence of $H_{\rm crit}$ for $t'=0$
and $\mu=0$. Continuous curve has been obtained for the PK model with 
$J=-1.56t$ ($\eta$--type pairing). The dashed line corresponds the
attractive pair hopping interaction $J=0.5t$. The dotted line
shows the critical field in the AH model with $U=t$.} 
\end{figure}

Fig. 3 shows the temperature dependence of $H_{\rm crit}$
obtained for attractive and for the repulsive pair
hopping interaction. These results are compared with
$H_{\rm crit}(T)$ calculated from the 2D AH model\cite{my}
with $U=-t$. We have adjusted the strength of the pair 
hopping interaction to obtain the same critical temperature
in the absence of magnetic field. For $J>0$ $H_{\rm crit}(T)$
in PK model is very close to that of AH model. It means
that in the case of $s$--wave pairing the Peierls factor in the
pair hopping term leads only  to a small decrease of superconducting
correlations. However, the temperature dependence of $H_{\rm crit}$
in the $\eta$--state differs qualitatively from the $s$--wave
case. Namely,  $H_{\rm crit}(T)$ has a very large slope for
a weak magnetic field and saturates already at relatively high temperature.
Such a behavior of the critical field resembles
$H_{pg}(T)$, that has recently been observed in ${\rm Bi_2 Sr_2 Ca
Cu_2 0_{8+y}}$.\cite{hpg} In Fig. 4 we compare our 
results and the experimental data.  

\begin{figure}
\vspace*{10mm}
\centering
\includegraphics[width=7.5cm]{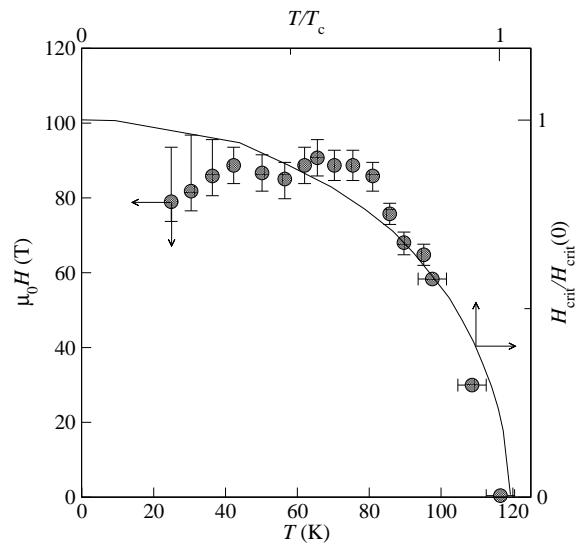}
\caption{
Fit of the theoretical results to the experimental data for the
pseudogap closing field $H_{pg}(T)$.\cite{hpg}
The continuous line represents $H_{\rm crit}(T)$ calculated
for the repulsive pair hopping interaction.
We have used the same model parameters as for the $\eta$--pairing in Fig. 3.}
\end{figure}
                                                                               
Within the Helfand--Werthammer
theory, the temperature dependence of critical field is predominantly
determined by the diamagnetic pair breaking mechanism. The Zeeman coupling
becomes important only for sufficiently strong magnetic field. This feature holds
also in the case of the lattice gas.\cite{maciek} In order to investigate the 
role of the Zeeman and orbital contributions in PK model, we have
repeated our calculations in the absence of the Zeeman term. The resulting
$H_{\rm crit}(T)$ is shown in Fig. 5. 

\begin{figure}
\centering
\includegraphics[width=7.5cm]{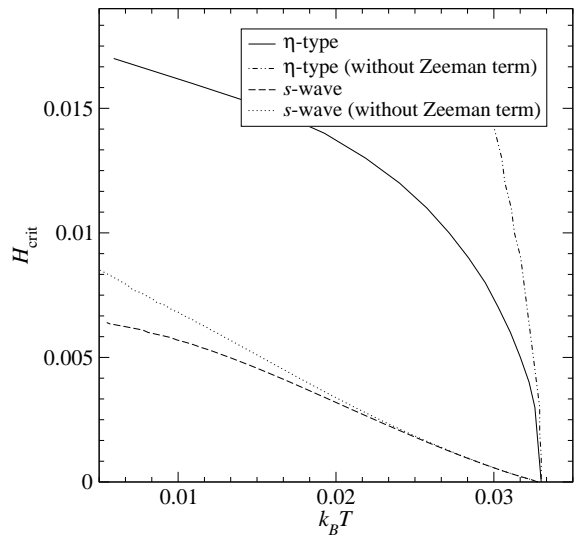}
\caption{$H_{\rm crit}(T)$ calculated for attractive and repulsive PK model with and without
the Zeeman coupling. We have used the same model parameters as in Fig. 3.}
\end{figure}

In contradistinction to the 
$s$--wave superconductivity, the diamagnetic pair breaking is of minor importance in the
case of $\eta$--pairing. This feature is responsible for extremely high values of $H_{\rm crit}$
in the absence of Zeeman term.
Experimental investigations\cite{hpg} show that the pseudogap closing field scales linearly with $T^*$. 
In Ref. \cite{hpg} the value of the scaling factor 
has been interpreted in favor of the Zeeman coupling as a mechanism that closes the pseudogap.
This pair breaking mechanism dominates also in the case of $\eta$--pairing. 


\section{Summary}

To summarize, we have investigated the PK model with
attractive as well as repulsive pair hopping interaction.
We have shown that the repulsive pair hopping term may 
lead to the occurrence of local minimum in the density of 
states, that is characteristic for pseudogap phase of underdoped 
cuprates. It originates from the spitting of the quasiparticle 
peaks. Despite the on--site pairing the magnitude of the splitting
is a direction--dependent quantity, provided that $t'\neq 0$. 
Anisotropy of the pseudogap is observed in ARPES 
experiments.\cite{arpes2} We have also calculated the temperature
dependence of $H_{\rm crit}$, defined as the highest magnetic field
for which there exists a non--zero solution for the order parameter.
We have found that in the case of $\eta$--type pairing $H_{\rm crit}(T)$
reproduces the experimental data for the pseudogap closing field.   
These features
do not occur for attractive pair hopping interaction. In this case 
the gap structure as  well as $H_{\rm crit}(T)$ are similar to those
obtained for AH model. 

Our approach to the critical field accounts both for Zeeman and 
diamagnetic pair breaking mechanisms. In the case of $s$--wave 
pairing inclusion of the Zeeman coupling does not lead to any 
essential changes in $H_{\rm crit}(T)$.  
On the other hand,
 Zeeman term is of crucial importance for $\eta$--pairing,
whereas the diamagnetic pair breaking is ineffective.
According to the experimental data the pseudogap is closed
by the Zeeman splitting.
  
 As we have previously\cite{my} shown, other models appropriate for short 
coherence superconductors have ground states (with $s$--wave or $d$--wave
symmetry) that are almost insensitive to the Zeeman interaction.
Therefore, the PK model with $J<0$ is unique in that the gap
is closed predominantly due to the Zeeman interaction.

Collecting the features: the presence of the pseudogap, its anisotropy,
Zeeman origin of $H_{\rm crit}$ (in agreement with the experimental data),
the presence of flux quantization and the Meissner effect,
(consistent with the preformed Cooper pairs scenario),
may lead to a tempting hypothesis
that the pair hopping can be responsible for the pseudogap. 
However, in order to avoid the problem of interpretation 
of the critical field, it should be verified beyond the mean--field 
level, discussed in this paper.
  
\acknowledgments
                                                                                
We acknowledge a stimulating discussion with S.~Robaszkiewicz. 
This work was supported in part by the Polish State Committee
for Scientific Research, Grant No. 2 P03B 050 23.

\end{document}